\documentclass[aps,twocolumn,floatfix,amsmath,amssymb,showpacs,showkeys]{revtex4}
 
\usepackage{amsmath}
\newcommand{\angstrom}{\text{\normalfont\AA}}
\usepackage[T1]{fontenc}
\usepackage{xcolor}
\newcommand{\be}{\begin{equation}}
\newcommand{\ee}{\end{equation}}
\newcommand{\ba}{\begin{align}}
\newcommand{\ea}{\end{align}}
\newcommand{\bn}{\begin{eqnarray}}
\newcommand{\en}{\end{eqnarray}}

\def\ba{\begin{eqnarray}}
\def\ea{\end{eqnarray}}

\def\({\left(}
\def\){\right)}
\def\g[{\left[}
\def\g]{\right]}
\def\g{{\left{}
\def\g}{\right}}

\usepackage{graphicx}
\usepackage[tight,TABTOPCAP]{subfigure}

\begin{document}


\title{Predicted antiferromagnetic-vortex dynamics driven by spin polarized current in thin discs}

\author{R. L. Silva}
\email{ricardo.l.silva@ufes.br}
\affiliation{Departamento de Ci\^encias Naturais, Universidade Federal do Esp\'{\i}rito Santo, 29932-540,  S\~ao Mateus, Esp\'{\i}rito Santo, Brazil.}

\author{R. C. Silva}
\email{rodrigo.c.silva@ufes.br}
\affiliation{Departamento de Ci\^encias Naturais, Universidade Federal do Esp\'{\i}rito Santo, 29932-540,  S\~ao Mateus, Esp\'{\i}rito Santo, Brazil.}



\author{R. J. C. Lopes}
\email{ricardo.lopes@ufv.br}
\affiliation{Departamento de F\'{i}sica, Universidade Federal de Vi\c{c}osa, 36570-900, Vi\c{c}osa, Minas Gerais, Brazil.}

\author{W. A. Moura-Melo}
\email{winder@ufv.br}
\homepage{https://sites.google.com/site/wamouramelo/}
\affiliation{Departamento de F\'{i}sica, Universidade Federal de Vi\c{c}osa, 36570-900, Vi\c{c}osa, Minas Gerais, Brazil.}

\author{A. R. Pereira}
\email{apereira@ufv.br}
\homepage{https://sites.google.com/site/quantumafra/home}
\affiliation{Departamento de F\'{i}sica, Universidade Federal de Vi\c{c}osa, 36570-900, Vi\c{c}osa, Minas Gerais, Brazil.}

\date{\today}
\begin{abstract}
We investigate vortex configuration in antiferromagnetic thin discs. It is shown that the vortex acquires oscillatory dynamics with well-defined amplitude and frequency which may be controlled on demand by an alternating spin polarized current. These findings may be useful for the emerging field of antiferromagnetic topological spintronics, once vortex dynamics may be controlled by purely electric means.
\end{abstract}
\maketitle

\section{Introduction and Motivation}
Topological excitations play an important role in many branches of physics, running from the low to the high energy regimes, from micro to macro scale structures. Currently, such excitations, like solitons, skyrmions and vortices, have attracted a great deal of efforts for they are claimed to be the key ingredients in novel mechanisms for spintronic devices \cite{topological-spintronic}.

In microsized ferromagnetic (FM) discs vortex-like patterns are observed as an in-plane close flux magnetization, minimizing the dipolar energy at the borders. This defines its chirality according to its flux is clockwise ($\mathcal{Q}=-1$) or counter-clockwise ($\mathcal{Q}=+1$). However, to minimize the exchange cost at the disc center, the magnetization revolves against the plane, in a small region of only a few exchange lengths, creating the vortex core with two distinct polarizations, up ($\mathcal{P} = +1$) or down ($\mathcal{P} = -1$). The pair ($\mathcal{Q},\mathcal{P}$) is known to govern many fundamental properties of vortex-state in FM micro- and nano-sized discs, such as its gyrotropic mode \cite{vortex-FM-girotropic}, switching of the vortex core (polarity reversal) \cite{vortex-FM-reversal,vortex-FM-reversal2} and vortex-pair excitation. A number of technological applications has been proposed as long as we may control one (or both) of these parameters on demand.

The scenario is quite distinct in antiferromagnetic (AFM) systems where a vortex-like excitation appears as a composed pattern of two anti-aligned sub-lattices, presumably yielding a much more rigid structure with no global dynamics, in clear contrast to what is verified for their FM counterparts. In addition, evidence of AFM vortex-like states is scarce whenever compared to FM vortices. To our knowledge, they were firstly observed by indirect ways, say, by inducing FM-ordered spins in AFM discs \cite{vortex-AFM-indireto,vortex-AFM-indireto1}. Later, direct observation of imprinted AFM vortex states in $CoO/Fe/Ag(001)$ thin microsized discs (an AFM layer of an AFM/FM bilayer system) has been reported \cite{vortex-AFM-direto}. In these systems, there appears a FM-AFM exchange coupling, in such a way that by measuring certain FM spin properties it provides a probe for their analogues lying in the antiferromagnet. In turn, once AFM systems are widely believed to play an important role in the emergent field of topological spintronics \cite{topological-spintronic,AFM-Spintronics,RMP-Baltz-2018-AFM-Spintronics}, further investigation upon AFM topological states is mandatory. Namely, under what condition an vortex emerge in an antiferromagnet and, whether and how it could acquire a controllable dynamics by external means are important questions to be answered. Along these lines, a systematic analysis concerning the dynamics of AFM solitonic-type topological excitations has recently appeared in Ref. \cite{DasguptaPRB95-2017}. Among other findings, authors claim that an AFM vortex may be put in motion provided that both magnetic field and spin current are applied. Although such an analytical procedure could capture a number of important aspects concerning such excitations in large and continuum systems, it remains to be investigated how their structure and dynamics will be affected in small and confined samples, where finite-size effects are crucial even for their appearance as stable configurations. Additionally, it should be mentioned that whenever compared to other AFM topologic-like configurations, like domain-walls and skyrmions, the literature dealing with AFM vortex-type excitations is still scarce \cite{RMP-Baltz-2018-AFM-Spintronics}.

Here, we show that the vortex configuration in a thin AFM disc may acquire dynamics only by applying spin current. More interesting, the vortex core is observed to oscillate with well-defined frequency and amplitude whenever a driven spin polarized alternating current (SPAC) is turned on. Its frequency and motion amplitude vary accordingly the frequency and strength of the current passing through the disc. This open us the possibility of practical control of AFM vortex dynamics by pure electric means. Additionally, as the vortex moves spin waves are emitted at the same frequency the magnetic dipoles oscillate,  similarly to what happens to their electric dipole counterparts. Therefore, by controlling the AFM vortex dynamics one rises a novel mechanism for generating spin waves with precise intensity and frequency with complete electric control, what may be relevant for AFM-based magnonic devices \cite{Cheng-PhysRevB98}.
\section{Methods}
Before presenting our model and its features, let us briefly discuss how vortex-type configurations may emerge in confined magnets. For that, let us consider an AFM sample with two N\'eel sublattices. Minimization of energy demands each moment of a sublattice must be counter-aligned to its nearest-neighbors of the other sublattice. In a continuum approach, where each dipole has magnetic momentum $M$ we take $M{\bf m}_1$ and $M{\bf m}_2$ to account for the magnetization corresponding to each sublattice. More effectively, one has the dominant staggered magnetization (N\'eel field), ${\bf n}=({\bf m}_1-{\bf m}_2)/2$, and the uniform field, $\boldsymbol\ell=({\bf m}_1+{\bf m}_2)/2$. Since $|{\bf m}_1|=|{\bf m}_2|=1$, at equilibrium, staggered and uniform fields are perpendicular each other everywhere, say, ${\bf n}\cdot \boldsymbol\ell=0$. In addition, one may use the equations of motion to cast out $\boldsymbol\ell$ in favor of the staggered field alone, which shall describes the magnetization states including its dynamics \cite{DasguptaPRB95-2017}.

In turn, the appearance of a specific configuration as the remanent ground-state in a finite-size magnet comes about as the balance between exchange and magnetostatic energies (say, for isotropic materials, in which anisotropic cost may be neglected). In these systems, effective magnetic charges are distributed in their volumes and surfaces, given by $\rho_m=-\nabla\cdot \bf{M}$ and $\sigma_m=\hat{n}\cdot \bf{M}$, where $\hat{n}$ is the unit vector normal to the borders at every point. [Indeed, in AFM systems such charges come about mainly due to fluctuations in the staggered magnetization, the N'eel field]. Therefore, to minimize the magnetostatic cost the magnetization tends to point parallel to the sample surface everywhere. However, as one goes towards the disc center, exchange energy density increases enormously (in a continuum framework, it is scaled with $r^{-2}$ , with $r$ being the distance from the disc center). Then, to regularize exchange cost, magnetic moments tend to revolve against the plane, developing out-of-plane projection, so that exactly at the center it is perpendicular to the disc face, as observed in confined discs comprising vortex-like patterns.\\

In turn, an AFM vortex-type state may be faced as a composite pattern with opposite polarity and chirality at the sublattices, in such a way that both parameters identically vanish for the whole AFM vortex configuration at equilibrium. This is the reason why its gyrotropic charge also falls off jeopardizing any dynamics as a whole, like the translational gyrotropic oscillation of the vortex core, observed in the ferromagnetic counterparts \cite{vortex-FM-girotropic,vortex-FM-reversal}. However, this cannot be understood as a vanishing dynamics of the AFM vortex at all. Indeed, it has been claimed that easy-axis AFM vortex, lying in a continuum and large system, may be put to move by a combined application of external magnetic field along with spin current \cite{DasguptaPRB95-2017}. In what follows, we shall argue that an AFM vortex comprised in a very thin disc may be driven to oscillate around its equilibrium position by solely applying a suitably spin current to the system. Our 'atomistic-type' model is summarized by the following Hamiltonian\footnote{As an atomistic-type model we also adopt units as it is usually done, for instance, $\hbar=c=1$ etc. However, for the sake of comparison, we quote SI units concerning our modeling. Typical AFM material exchange constant goes around $J\sim 10^1-10^2 {\rm meV}\sim 10^{-21} {\rm Joules}$. A dipole/spin at site $i$, $\vec{m}_i= \mu_{\rm B}\vec{\mu}_i$, so that $\vec{\mu}_i$ is dimensioless ($ \mu_{\rm B}$ is the Borh magneton). Frequency, $\omega$, is measured in units of $J/\hbar \sim 10^{12}-10^{13} {\rm Hz}$, while spin current density magnitude, $\jmath_0$, is given in terms of   ${\rm charge}\times{\rm frequency}/{\rm area} \sim 10^{12}-10^{13} {\rm A/m^2}$.}:
\begin{equation}
H= +J\sum_{\langle i,j\rangle} \vec{\mu}_i \cdot \vec{\mu}_j  - \sum_{\alpha=1,2} \sum_k \lambda_\alpha (\vec{\mu}_k \cdot \hat{n}_{k,\alpha})^2\,. \label{H-model}
\end{equation}
In the first term, the sum runs over nearest-neighbors (NN) magnetic dipoles, $\vec{\mu}$, and it accounts for the usual exchange energy, favoring the anti-alignment of dipoles since $J>0$. The second term mimics the role of effective magnetic charges at the borders ('magnetostatic-like cost', see discussion below). For $\alpha=1$ the sum over $k$ is taken along the entire disc face while for $\alpha=2$ such a sum is restricted on the disc radius, say, sites where their dipoles lies at a distance $R$ from the disc center. If we consider the face of the disc lying on the $xy$-plane, then the normal unity vectors read $\hat{n}_{k,1}=\hat{z}, \forall k$ while $\hat{n}_{k,2}=\hat{r}$, with $\hat{r}=\vec{r}/|\vec{r}|= (x\hat{x} + y\hat{y})/|\vec{r}|$. It should be emphasized that Hamiltonian (\ref{H-model}) has been successfully applied to study vortex structure in ferromagnetic discs \cite{Ricardo2008}.\\

In addition, it is augmented by simulated annealing, which yields the dipoles to an equilibrium configuration at a given finite temperature, and eventually spin dynamics to relaxing the system down to zero temperature, when the dipoles should presumably remain stationary. Although such a condition is expected to occur at the 'bulk' (say, the internal part of the thin disc) perhaps the same does not takes place at the borders, where non-trivial boundary conditions come about. Indeed, if the dipoles at the borders cannot reach the static equilibrium, then their fluctuations there could account for the 'artificial' magnetostatic cost inserted in the Hamiltonian above\footnote{Actually, AFM spins fluctuations naturally occur, even at zero temperature, for instance, if spins are disposed at the corners of a triangle. By virtue of the underlying arrangement, their mutual nearest-neighbor interactions cannot completely minimize the total energy, and they are said to be geometrically frustrated\cite{Geometrical-frustration}. In words, AFM spins lying on a triangular lattice cannot be completely frozen. We may wonder whether a similar scenario is taking place with the AFM dipoles at the thin disc border. Further experimental and theoretical studies are demanded to shed light upon such a question. This could also answer if confined AFM vortices appear stationary or dynamically stable, at all.}.
\begin{figure}[h!]
	\centering
	\subfigure[]{\includegraphics[width=40.0mm]{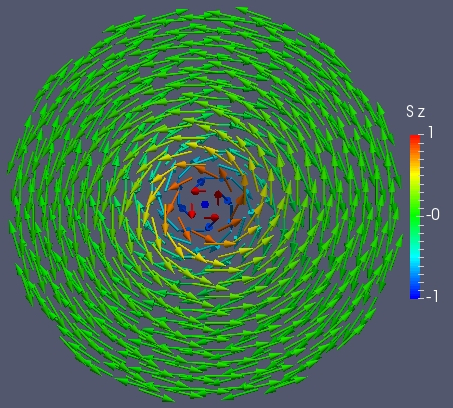}}
	\subfigure[]{\includegraphics[width=40.0mm]{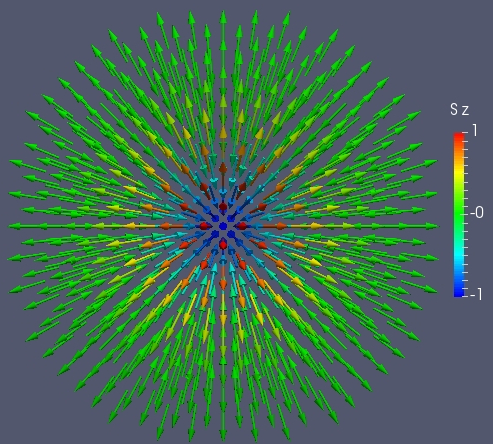}}
	\vskip 0.1cm
	\subfigure[]{\includegraphics[width=40.0mm]{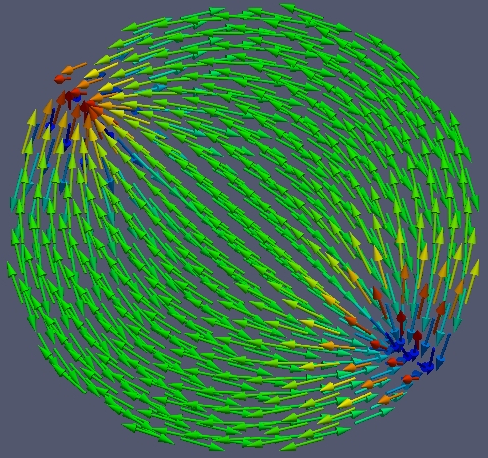}}
	\subfigure[]{\includegraphics[width=40.0mm]{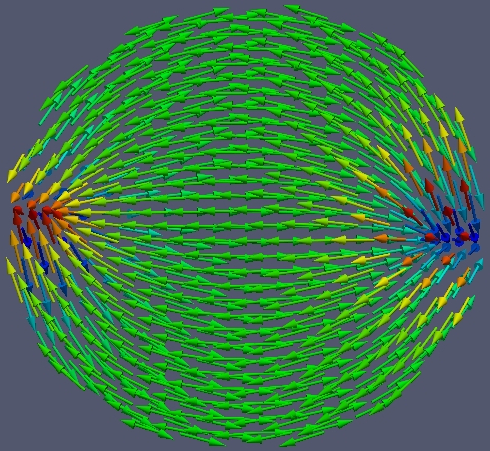}}
	\caption{(Color online) Curly- and divergent-type AFM vortex profiles, panels (a) and (b), respectively.Both patterns have been observed in very thin AFM microsized discs, reported in Ref. \cite{vortex-AFM-direto}. Intermediate states, known as onion-like configurations, show up interpolating curly and divergent vortices. Two of these patterns are shown in panels (c) and (d), obtained with $\lambda_2/J=+0.4$ and $\lambda_2/J=+0.6$, respectively. Such states have not been observed in AFM systems by now, although their FM analogues have shown up in certain confined geometries.}
	\label{fig1}
\end{figure}
By adjusting parameters $J$ and $\lambda_\alpha$ one gives rise to both curly-type and divergent-like vortices, which is an important feature of the model above, once such sorts of distinct vortices do appear in thin AFM discs. More precisely, FM-AFM bilayers of single-crystalline $NiO/Fe/Ag$ and $CoO/Fe/Ag$ 2-$\mu$m-diameter discs have been observed to support curly-type AFM vortex whenever the oxide layer is very thin, say, $\sim 0.6 {\rm nm}$, while a divergent vortex takes place if the thickness is increased to around $3{\rm nm}$ \cite{vortex-AFM-direto}. Figure \ref{fig1} display their profiles: while in a curly vortex the dipoles tend to be parallel to the disc border, at $r=R$, divergent vortex dipoles appear to point in/outward this border. In our simulations we have fixed $\lambda_1/J = +0.05$, so that curly vortex  has emerged as the ground-state of very thin AFM disc whenever $\lambda_2/J \ge +0.8$, while divergent-like pattern is favored whenever $\lambda_2/J \le -0.8$. For intermediate values, $-0.8<\lambda_2/J<+0.8$ onion-type states also show up. Such states resembles a pair of half-vortex each of them comprised along a half disc, see Fig. \ref{fig1}. [We may wonder whether such states could be observed in similar systems as those reported in ref.\cite{vortex-AFM-direto}, for instance, $CoO/Fe/Ag$ 2-$\mu$m-diameter discs but with thickness around $1-2 {\rm nm}$]. In addition, they are the predominant ground-state whenever $-0.35<\lambda_2/J<+0.35$. The vortex configuration so emerged may be viewed as a composite structure of a pair of sub-vortices each of them associated with one AFM sub-lattice. Each sub-vortex has a pattern quite similar to its counterpart in a FM disc, but they always appear with opposite polarity and chirality each other, yielding net vanishing of such quantities for the whole AFM vortex at equilibrium.\\

After having stabilized the vortex configuration in the thin AFM disc, we depart to study its dynamical properties. For that we employ Landau-Lifshitz-Gilbert-like equation augmented with the spin current torque on the dipoles (Berger-Slonczewski term), as below:

\begin{eqnarray}
& \frac{d}{dt} \vec{\mu}_i=& -\gamma\vec{\mu}_i \times H_{\rm eff} - \alpha\vec{\mu}_i \times \frac{d}{dt}\vec{\mu}_i + \vec{\tau}_i\,,\\ 
\nonumber\\
& & \vec{\tau}_i= p \frac{\nu_{\rm cell}}{2e} \Big[\,(\vec{\jmath}\cdot \nabla) \vec{\mu}_i + \beta\, \vec{\mu}_i \times (\vec{\jmath}\cdot \nabla) \vec{\mu}_i \, \Big], \hskip 0.5cm
\end{eqnarray}
where $H_{\rm eff}=-\frac{1}{\mu}\frac{d}{d\vec{\mu}_i}H$ is the effective local field acting on dipole $\vec{\mu}_i$ while $\alpha$ is the Gilbert damping parameter (we have set $\alpha=0.1$ in our simulations). $\vec{\tau}_i$ is the spin transfer torque applied on $\vec{\mu}_i$, where
$p$ is the spin current polarization ($p=+1$ for {\em up} while $p=-1$ for {\em down} electrons; we have taken $p=-1$ in our calculations); $\nu_{\rm cell}=a^3$ is the unity cell volume of the material ($a\sim\angstrom$, is the cell lattice spacing); $e>0$ is the magnitude of the fundamental charge; $\beta$ is the non-adiabaticy factor (we have fixed $\beta=0$, once no spin torque damping has been considered in our study). Note that no external magnetic field is applied to the AFM vortex lying on the thin disc.\\

\begin{figure}[h!]
	\center
	\includegraphics[width=7.5cm]{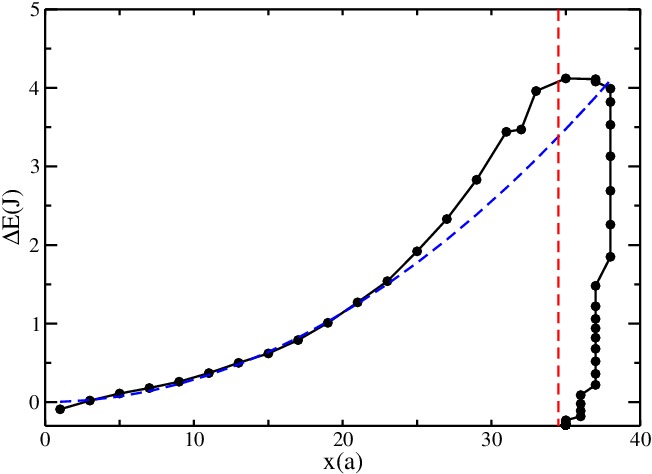} 
	\caption{(Color online) Shows the AFM vortex energy (relative to its energy at the disc center as function of its core position along $x$. For small displacements around the equilibrium position the energy goes with $x^2$ (blue curve), but deviates from such a behavior far from the disc center. Whenever the spin current torque pushes the vortex core against the disc border, it is eventually destroyed and its energy falls off rapidly, marked by the red traced vertical line at $x\approx35 a$. Here, we have taken $\jmath_0=0.2$ and disc radius $R=40a$.}
	\label{vortex-enery}
\end{figure}

Firstly, a spin polarized direct current (SPDC) is applied to the disc along the $x$-axis, $\vec{\jmath}=\jmath_0 \hat{x}$. The resulting displacement of the vortex core is plotted against its energy in Fig. \ref{vortex-enery}. As may be realized, its energy increases with $x^2$ for relatively small displacements, and departs from this harmonic behavior for large $x$'s. Whenever the vortex core is pushed against the border it is eventually destroyed and its energy quickly fall off.
\begin{figure}[h!]
	\centering
	\subfigure[$t=0.0$]{\includegraphics[width=40.0mm]{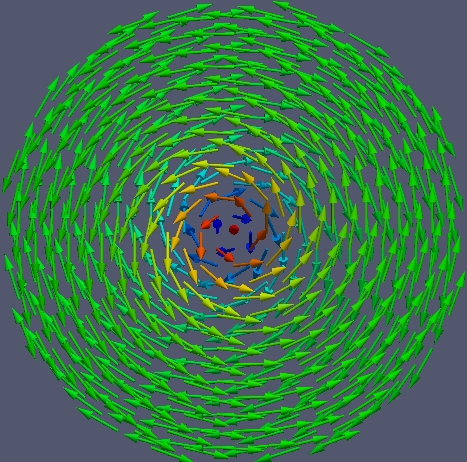}}
	\subfigure[$t=T/4$]{\includegraphics[width=40.0mm]{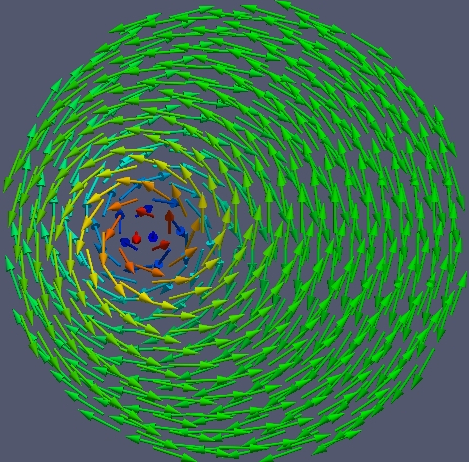}}
	\vskip 0.1cm
	\subfigure[$t=T/2$]{\includegraphics[width=40.0mm]{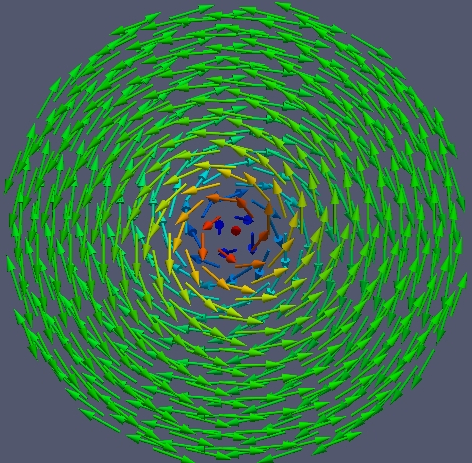}}
	\subfigure[$t=3T/4$]{\includegraphics[width=40.0mm]{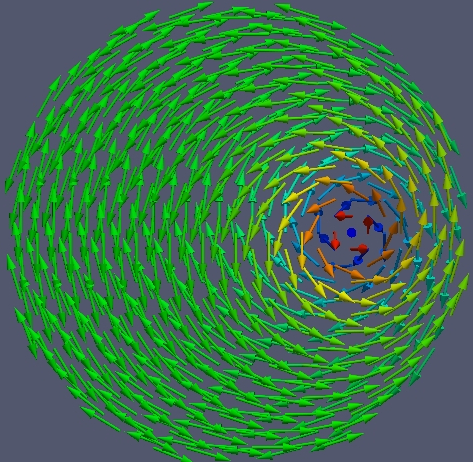}}
	\vskip .4cm
	\includegraphics[width=7.5cm]{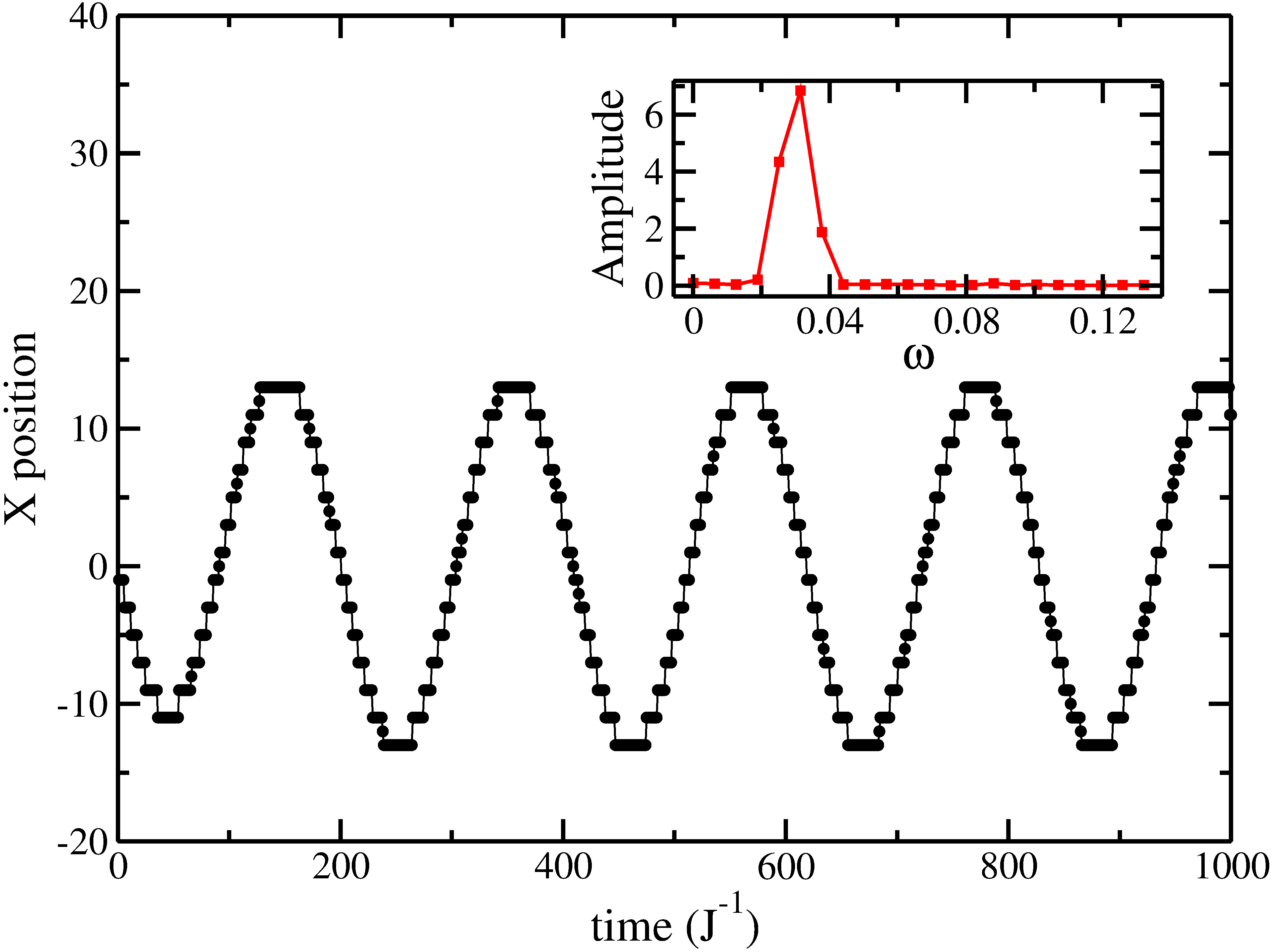} \hskip0.5cm 
	\caption{\label{Vortex-oscillation}(Color online) The harmonic oscillatory dynamics of the AFM vortex core driven by SPAC applied along $x$-axis (horizontal direction). Panels (a)-(d) show the vortex position at different time while its core displacing along $x$-axis ($T$ is the period of a complete cycle). At the bottom panel it is plotted vortex core position as function of time for a number of cycles, clearly showing that the driven vortex dynamics is kept while the current is applied. The inset shows the amplitude of the oscillation against the SPAC frequency, $\omega$. Notice that the core oscillates at the same frequency as the current one, $\omega_{\rm vortex}=\omega=0.03 J$.}
\end{figure}
\section{Results and Discussion}
\begin{figure}[h!]
	\centering
	\subfigure[$t=T/4$]{\includegraphics[width=40.0mm]{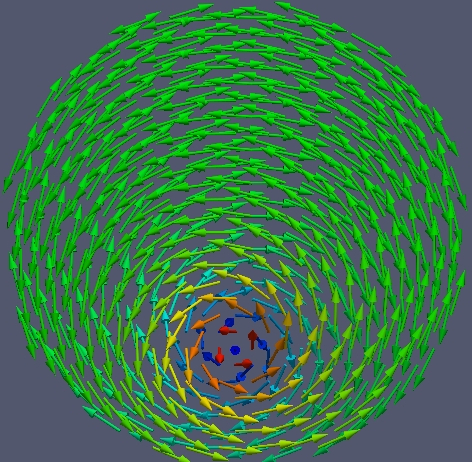}}
	\subfigure[$t=T/2$]{\includegraphics[width=40.0mm]{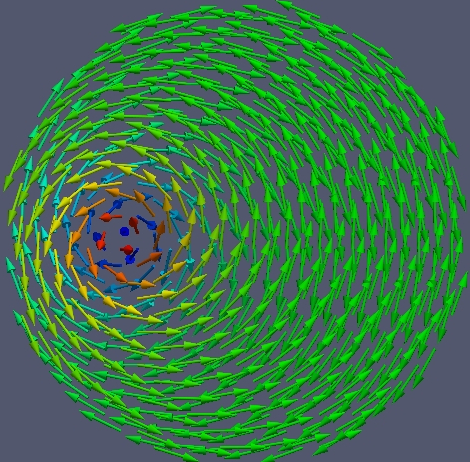}}
	\vskip 0.1cm
	\subfigure[$t=3T/4$]{\includegraphics[width=40.0mm]{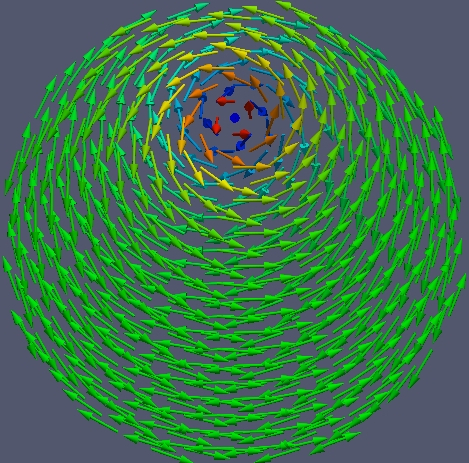}}
	\subfigure[$t=T$]{\includegraphics[width=40.0mm]{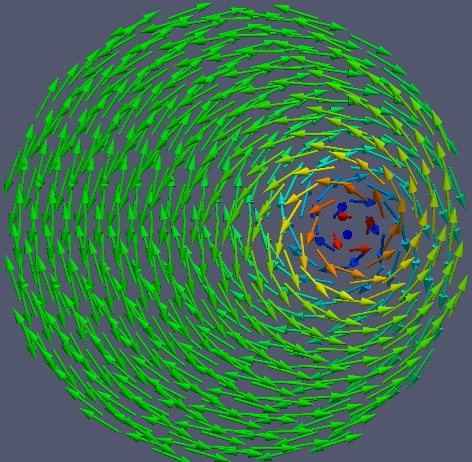}}
	\vskip .4cm
	\includegraphics[width=7.5cm]{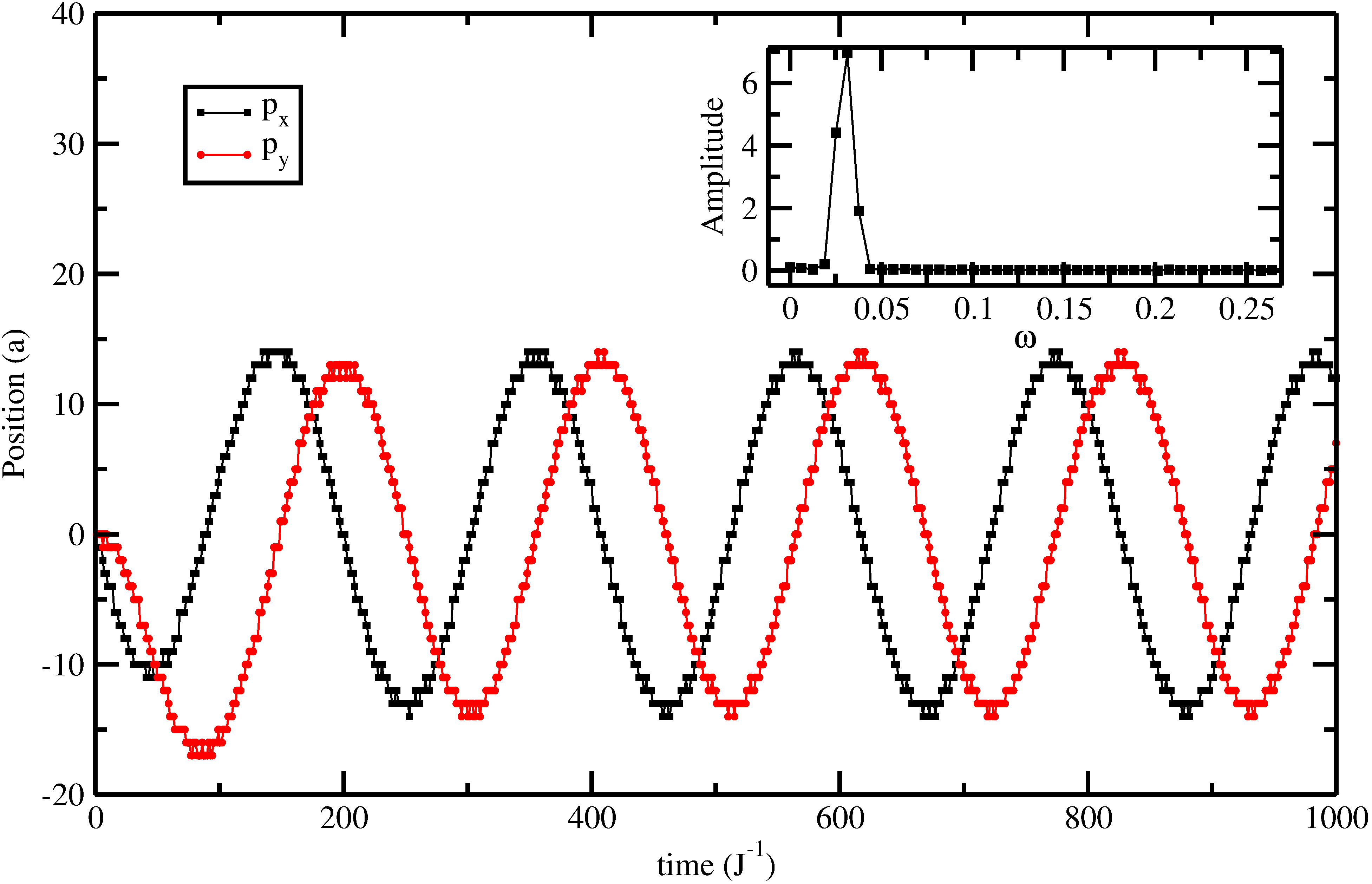}
	\caption{\label{Vortex-oscillation-XY} (Color online) The circular oscillatory dynamics of the AFM vortex core driven by two SPAC's applied along $x$ and $y$ directions. Panels (a)-(d) show the vortex position at different times ($T$ is the period of a complete cycle; for $T=0$ the vortex position and pattern are the same as in Fig. \ref{Vortex-oscillation}). At the bottom panel the vortex core position is plotted as function of time for several cycles. The inset shows the amplitude of the oscillation against the SPAC frequency, $\omega$. The core oscillates at the same frequency as the applied current one, $\omega_{\rm vortex}=\omega=0.03 J$.}
\end{figure}
\vskip .1cm
Now, we depart to study the AFM vortex dynamics. For that, a spin polarized alternating current (SPAC), $\vec{\jmath}=\jmath_0 \cos(\omega t) \hat{x}$, is then turned on and it trespasses the disc along $x$. For numerical purposes, we have set $p=-1$ and we have used $\jmath_0= 0.2$ and $\omega=0.03$. This causes the AFM vortex pattern to oscillate along the current direction at this same frequency, $\omega_{\rm vortex}=\omega$. The vortex core amplitude is controlled by suitable choice of the current strength, $\jmath_0$. Fig. \ref{Vortex-oscillation} shows a number of frames of the vortex oscillation along with its position against time. A better visualization of the vortex dynamics itself may be found in Movie 1 (Supplementary Material).

Once the vortex motion is driven by the SPAC, we may apply a second current, along $y$, $\vec{\jmath}_y=\jmath_0 \sin(\omega t)\hat{y}$, in order to produce a circular oscillation of the vortex core, resembling the well-known girotropic motion observed for FM vortex (see discussion below). Fig. \ref{Vortex-oscillation-XY} shows a number of frames of the AFM vortex circular motion (Movie 2, Supplementary Material, presents its complete dynamics in more details). We should remark that the present AFM vortex oscillation driven by SPAC should not be confused with the girotropic dynamics observed for vortex-type excitations lying on FM discs and similar geometries. This is a mechanic-like motion which can be induced by several means, including magnetic field and spin torque effects\cite{GuslienkoPRL2008}. Namely, girotropic frequency depends upon the material and geometry of the sample and it ceases slowly by damping effects. AFM vortex oscillation driven by alternating current is a more rigid dynamics which ceases abruptly whenever the SPAC is turned off.\\

That we have a genuine vortex driven dynamics induced by an alternating current may be also realized by adopting a collective coordinate approach, $b_i=(b_x,b_y)$, like that presented in the work of Ref.\cite{TetriakovPRL2013}. Indeed, taking this approach to the AFM vortex core, we obtain\footnote{We have adapted the main result from Ref.\cite{TetriakovPRL2013}, Eq. (5), to our case by: i) assuming that the vortex motion as a whole is equivalent to its core dynamics. This simplifies the problem of the whole extended vortex pattern dynamics to that of a particle alone. This is strictly valid if the vortex pattern is not deformed so much. In our case, this is ensured if we restrict the vortex dynamics to relatively small oscillations. Thus, the inertia tensor $M^{ij}$ reduces to a unique component, $m$; ii) noticing that the AFM vortex core oscillates under SPAC practically without dissipation, once its amplitude remains essentially unaltered over several oscillations, as shown in Figs. \ref{Vortex-oscillation} and \ref{Vortex-oscillation-XY}. Thus the friction term, proportional to $\dot{b}_j=db_j/dt$ in Eq. (5), Ref.\cite{TetriakovPRL2013}, may be neglected for our purposes. Finally, we are left with our Eq. (\ref{collective-equation}).}:
\begin{equation}\label{collective-equation}
m \ddot{b}_i=F_i\,,
\end{equation} 
where $F_i=(F_x, F_y)$ are the components of the effective force acting on the vortex core. Such a force is proportional to the applied SPAC, so that the core dynamics is identical to that of a harmonic oscillator system lying on the $xy$ plane.\\
\section{Concluding remarks}
In summary, we have shown that vortex-type configuration lying on antiferromagnetic thin discs may be put in motion only by applying a spin polarized current throughout the sample. More specifically, we have realized that the application of a spin polarized alternating current drive the vortex dynamics to oscillate in a controlled way. Actually, vortex core amplitude and frequency may be adjusted on demand just by tunning the strength and frequency of the applied current. Therefore, despite the widely belief of a rigid structure of antiferromagnetic vortices, our findings not only put forward the possibility of vortex dynamics, they have also shown that such a motion may be completely controlled by purely electric means. Additionally, AFM vortex oscillation may be used to produce spin waves (magnons) with well-defined frequency and intensity, making the present system also relevant to magnonic-based mechanisms, as recently raised in Ref.\cite{Cheng-PhysRevB98}. Finally, we claim that our findings may be useful for the emergent field of antiferromagnetic spintronics, which is widely based upon the utilization of topological-like excitations as the key ingredients for novel spintronic mechanisms.\\
\section{Acknowledgments}
The authors thank CAPES, CNPq and FAPEMIG (Brazilian agencies) for partial financial support.

\end{document}